\documentclass[a4paper,11pt,twoside]{article}
 \pdfoutput=1
\usepackage[english]{babel}
\makeatletter
\usepackage{geometry,graphicx,color}
\usepackage{amsmath,amssymb}
\usepackage[pdftex]{hyperref}
\usepackage{enumerate}
\numberwithin{equation}{section}
\newcommand\bbone{{\mathbb{I}}}

\newcommand{\institute}[1]{\newcommand{\@institute}{#1}}
\renewcommand{\maketitle}{
\vspace*{0.5\baselineskip}
{% title
\center\LARGE\noindent\@title\par
}%
\vspace{1.5\baselineskip}
{% author
\center\normalsize\noindent\ignorespaces\@author\par
}%
\vspace{0.5\baselineskip}
{% institute
\center\normalsize\ignorespaces\@institute\par
}%
\vspace{2\baselineskip}
}%
\let\OLDthebibliography\thebibliography%
\renewcommand\thebibliography[1]{%
\OLDthebibliography{#1}%
\setlength{\parskip}{0pt}%
\setlength{\itemsep}{0pt plus 0.3ex}%
}%
\begin{document}
\title{Gauge theories on $\kappa$-Minkowski spaces: \\Twist and modular operators}
\author{Philippe Mathieu$^a$, Jean-Christophe Wallet$^b$}
\institute{%
\textit{$^a$Department of Mathematics, 
University of Notre Dame\\Notre Dame, IN46556, USA.\\
$^b$IJCLab, CNRS, 
University Paris-Saclay, 91405 Orsay, France
}\\%
e-mail: {\texttt{pmathieu@nd.edu}}, {\texttt{jean-christophe.wallet@th.u-psud.fr}}\\[1ex]%
}%
\maketitle
%-- ABSTRACT --------------------------------------------------------------------------%
\begin{abstract} 
We discuss the construction of $\kappa$-Poincar\'e  invariant actions for gauge theories on $\kappa$-Minkowski spaces. We consider various classes of untwisted and (bi)twisted differential calculi.  Starting from a natural class of noncommutative differential calculi based on a particular type of twisted derivations belonging to the algebra of deformed translations, combined with a twisted extension of the notion of connection, we prove an algebraic relation between the various twists and the classical dimension d of the $\kappa$-Minkowski space(-time) ensuring the gauge invariance of the candidate actions for gauge theories. We show that within a natural differential calculus based on a distinguished set of twisted derivations, d=5 is the unique value for the classical dimension at which the gauge action supports both the gauge invariance and the $\kappa$-Poincar\'e invariance. Within standard (untwisted) differential calculi, we show that the full gauge invariance cannot be achieved, although an invariance under a group of transformations constrained by the modular (Tomita) operator stemming from the $\kappa$-Poincar\'e invariance still holds.

\end{abstract}
\newpage
%============================================================================%

\section{Introduction}

A rather common belief is that the usual notion of space-time modelised by a smooth continuous manifold becomes likely unsuitable near the Planck scale to reconcile quantum mechanics and gravity. Noncommutative Geometry offers an appealing way to deal with this long standing problem \cite{Doplich1} and basically amounts to replace the classical structures linked to the manifold by their noncommutative (quantum) counterparts. Among the related quantum (noncommutative) spaces considered so far, the $\kappa$-Minkowski space is often regarded as a promising candidate for a quantum space-time underlying the description of quantum gravity possibly in some regime/limit. The $\kappa$-Minkowski space can be conveniently viewed, in a first (algebraic) stage, as the enveloping algebra of the solvable Lie algebra defined by $[x_0,x_i]=\frac{i}{\kappa}x_i,\ \ [x_i,x_j]=0,\ \ i,j=1,\cdots, (d-1)$. Here, $x_0,\ x_i$, the ``noncommutative coordinates'', are self-adjoint operators and $\kappa$, a positive real number, is the deformation parameter with the dimension of a mass, which is assumed to be of the order of the Planck mass. \\

The characterization of the $\kappa$-Minkowski space has been carried out in \cite{majid-ruegg} using the Hopf algebra bicrossproduct structure of the $\kappa$-Poincar\'e quantum algebra $\mathcal{P}_\kappa $\cite{luk1} whose coaction on $\kappa$-Minkowski is covariant, as being the dual of a subalgebra of $\mathcal{P}_\kappa $ often called the ``algebra of deformed translations''. Hence, $\mathcal{P}_\kappa $ may be viewed as coding the quantum symmetries of the $\kappa$-Minkowski space.
Many works in the literature deal with related algebraic structures, encompassing quantum group viewpoint \cite{leningrad} as well as twist deformations. The review \cite{luk2} on these aspects  covers most of all the relevant references. Besides, possible phenomenological implications of these structures have attracted a lot of interest for a long time, resulting in numerous works, in particular on Doubly Special Relativity, modified dispersion relations as well as relative locality \cite{ame-ca1,reloc}. \\

These features have triggered for more than two decades a great interest for Noncommutative Field Theories (NCFT) on $\kappa$-Minkowski spaces; see for instance  \cite{habsb-impbis}-\cite{hrvat-1}. This interest was even increased by the observation \cite{maatz} that the integration over the gravitational degrees of freedom in the (2+1)-d quantum gravity with matter gives rise to a NCFT invariant under a kappa deformation of the Poincar\'e algebra with non-trivial action on the multi-particle states. This observation, albeit valid only in (2+1)-d, reinforces the idea that $\kappa$-Minkowski and $\kappa$-Poincar\'e structures may well be of relevance in a description of the (3+1)-d quantum gravity.  Contrary to NCFT on the popular noncommutative spaces, such as Moyal spaces or $\mathbb{R}^3_\lambda$, for which many results on (perturbative) quantum and renormalisability properties have been obtained, see e.g. \cite{Grosse-1}-\cite{vign-sym}, it appears that the corresponding quantum properties for NCFT on $\kappa$-Minkowski spaces have been poorly explored until recently \cite{mercati1}, \cite{PW2018}, \cite{PW2018bis}.\bigskip

In \cite{PW2018}, we introduced a star-product for $\kappa$-Minkowski spaces from which a systematic analysis of the quantum properties of NCFT on these spaces is now possible. Note that this product was derived in \cite{DS} in a somewhat different context. This product is simply obtained from a combination of the Weyl-Wigner quantization map with the convolution algebra{\footnote{See \cite{dana} for mathematical details.}} of the affine group $\mathbb{R}\ltimes\mathbb{R}^{(d-1)}$.  It is nothing but an extension of the construction of the star product used in \cite{wal-16}, \cite{poul-wal} for the case of the noncommutative space $\mathbb{R}^3_\lambda$, stemming from standard properties of harmonic analysis on the $SU(2)$ Lie group. The $\kappa$-Poincar\'e invariance is a reasonable requirement for a NCFT to be of potential interest for Physics, in particular to a regime near the Planck scale, in view of the close interplay between $\kappa$-Poincar\'e  and $\kappa$-Minkowski algebras, which might represent a part of a quantum version of the present description of the so far experimentally accessible physics. As pointed out in \cite{PW2018}, \cite{DS}, the $\kappa$-Poincar\'e invariance of an action is obtained whenever it involves the simple Lebesgue integral which however behaves as a twisted trace w.r.t. the star product. But this implies that the action is rigidly linked with a KMS weight \cite{kuster} together with a distinguished one-(real)parameter group of automorphisms, called the (Tomita) group of modular automorphisms \cite{takesaki}. Hence, cyclicity of the trace is traded for the above KMS property while the generator of the group of modular automorphisms, called the modular operator, is nothing but the the modular twist defining the above trace. Possible physical consequences in connection with \cite{ConRove} where discussed in \cite{PW2018}.  The one-loop properties of various families of $\kappa$-Poincar\'e invariant (complex) scalar NCFT on the four dimensional $\kappa$-Minkowski whose commutative limit is the usual massive $\phi^4$ theory in \cite{PW2018}, \cite{PW2018bis}. It is shown that the perturbative UV behavior for the 2- and 4-point functions is essentially controlled by the contribution of the modular twist while IR singularities (in the 2-point functions) that would signal occurrence of perturbative UV/IR mixing do not appear in many considered NCFT. A particular class of NCFT on 4-d $\kappa$-Minkowski space, called orientable NCFT {\footnote{See e.g. first of ref. \cite{vign-sym} for the first use of this terminology in the Moyal spaces case.}}, has been closely examined in \cite{PW2018bis} and found to have in particular (one-loop) scale-invariant couplings, signaling the vanishing of the beta functions at one-loop. \\

In this paper, we construct $\kappa$-Poincar\'e  invariant actions for gauge theories defined on $\kappa$-Minkowski spaces. The problem of constructing $\kappa$-Poincar\'e  invariant actions for gauge theories on $\kappa$-Minkowski spaces is not an easy task at least because the trace is no longer cyclic. As a result, a twist appears upon cyclic permutation of the factors which prevents the various factors arising from gauge transformations to balance each other.  We first consider untwisted noncommutative differential calculi such as those usually considered in the physics literature and show that no polynomial action in the curvature with reasonable commutative limit can have a full gauge invariance; these actions however remain invariant under a group of transformations constrained by the modular operator which may strongly limit the usefulness of this invariance. We then consider a natural class of noncommutative differential calculi based on a particular type of twisted derivations belonging to the algebra of deformed translations, combined with a twisted extension of the notion of connection. The main result of this paper can be summarized as follows: \\
We prove an algebraic relation between the various twists and the classical dimension of the $\kappa$-Minkowski space which ensures the gauge invariance of the candidate actions for gauge theories. Fixing the twists fixes the unique value of the dimension at which the gauge invariance can be achieved while fixing the dimension severely restricts the allowed twists. We then select a natural differential calculus based on a distinguished set of twisted derivations leading in particular to a Dirac operator with required properties to be used in a (twisted) spectral triple \cite{como-1} for $\kappa$-Minkowski space \cite{matas}. Within this framework, we show that d=5 is the unique value for the classical dimension of $\kappa$-Minkowski space at which the gauge action supports both the gauge invariance and the $\kappa$-Poincar\'e invariance. \\

The paper is organized as follows. The section \ref{section2} collects the useful properties of the star product for $\kappa$-Minkowski spaces and recall the main consequences of the $\kappa$-Poincar\'e invariance requirement, in particular properties of the Lebesgue integral as a twisted trace w.r.t. the star product, related to a KMS weight. In the section \ref{utomita}, we consider the standard situation of untwisted differential calculi. In the subsection \ref{section31}, we extend a derivation-based differential calculus to a (noncommutative) differential calculus based on twisted and bitwisted derivations which are natural extensions of derivations in the present context. In the subsection \ref{section32}, we consider twisted derivations and related differential calculi  and extend the notion of noncommutative connection to twisted connections together with their gauge transformations. The subsections \ref{section43} and \ref{section44} deal with the gauge invariance of actions, where we exhibit a relation between the various twists, the classical dimension of the $\kappa$-Minkowski space and the gauge invariance. We finally discuss the results. \\

\paragraph{Notations:} In the following, we denote spacelike (resp. timelike) coordinates by Latin indices $i,j,\hdots=1,2,...,(d-1)$ (resp. $0$ indices) where d denotes the classical dimension of the $\kappa$-Minkowski space, Greek indices ($\mu,\nu,\hdots$) run from $0$ to $(d-1)$. For any d-vector $x$, $x:=(x_\mu)=(x_0,\vec{x})$ and $x.y:=x_\mu y^\mu=x_0y_0+\vec{x}\vec{y}$ (we work with Euclidean signature). Einstein summation convention for repeated indices is understood. The Fourier transform of $f\in L^1(\mathbb{R}^d)$ is given by $(\mathcal{F}f)(p):=\int d^dx\ e^{-i(p_0x_0+\vec{p}.\vec{x})}f(x)$ with inverse $\mathcal{F}^{-1}$ and $\bar{f}$ its complex conjugate. $\mathcal{S}_c$ is the space of Schwartz functions on $\mathbb{R}^4=\mathbb{R}\times\mathbb{R}^{(d-1)}$ with compact support in the first variable. 

\section{Star product for $\kappa$-Minkowski space and twisted trace.}\label{section2}

In this section, we first recall the main properties of the star product used to model the $\kappa$-deformation of the Minkowski space stemming from a Weyl-Wigner quantization. This is summarized in the subsection \ref{section21} as well as in the appendix \ref{apendmoyal} together with the main properties of the algebra modeling the $\kappa$-Minkowski space. In the subsection \ref{kappatwist}, the essential role played by the Lebesgue integral in the construction of $\kappa$-Poincar\'e invariant action functionals is outlined. Insisting on the $\kappa$-Poincar\'e invariance necessarily implies that these action functionals are KMS weight, hence implying the appearance of a KMS condition on the algebra of fields. This is (equivalently) reflected by the fact that the Lebesgue integral defines a twisted trace with respect to the star product, whose twist operator is linked to the Tomita operator generating the modular group of $^*$-automorphisms of the KMS structure \cite{PW2018}. \\

\subsection{Basic properties of the star product for $\kappa$-Minkowski spaces}\label{section21}
As explained in \cite{PW2018, DS}, a star product defining the $\kappa$-Minkowski space can be obtained from a straightforward adaptation of the Weyl-Wigner quantization scheme leading to the popular Moyal product. The corresponding main steps underlying the construction are outlined in the appendix \ref{apendmoyal} for the sake of completeness.\\
 As far as the construction of a star product defining the 4-dimensional $\kappa$-Minkowski space is concerned, one starts from the convolution algebra $L^1(\mathcal{G})$, the space of integrable $\mathbb{C}$-valued functions on the non unimodular affine group
\begin{equation}
\mathcal{G}:=\mathbb{R}^+_{/0}\ltimes_\phi\mathbb{R}^3,\label{crossaffine}
\end{equation}
where $\phi:\mathbb{R}^+_{/0}\to\textrm{Aut}(\mathbb{R}^3)$ is the adjoint action of $\mathbb{R}^+_{/0}$ on $\mathbb{R}^3$. Here, the convolution product $\circ$ is defined w.r.t. to the right-invariant Haar measure which simply reduces to the Lebesgue measure. Its appearance in the action functionals for NCFT insures the $\kappa$-Poincar\'e invariance which seems to be a physically reasonable requirement \cite{PW2018, DS}. Now, for any $t\in\mathcal{G}$ and $f, g\in{L^1(\mathcal{G})}$, the convolution product is given by 
\begin{equation}
(f\circ g)(t):=\int_{\mathcal{G}}d\nu(s)\ f(ts^{-1})g(s). 
\end{equation}
The involution turning $L^1(\mathcal{G})$ into a $^*$-algebra is given by 
\begin{equation}
f^*(t):=\bar{f}(t^{-1})\Delta_{\mathcal{G}}(t),
\end{equation}
in which $\Delta_{\mathcal{G}}$ is the modular function relating right- and left-invariant Haar measures. For mathematical details, see e.g. \cite{dana}. Then, identifying functions in ${L^1(\mathcal{G})}$ as functions on the momentum space,  as it is done for the Moyal case (see appendix \ref{apendmoyal}), namely
\begin{equation}
f(t)\ \to \mathcal{F}f(p_0,\vec{p})\label{param-conseq1}
\end{equation}
while one has
\begin{equation}
\Delta_{\mathcal{G}}(t)\ \to \Delta_{\mathcal{G}}(p_0,\vec{p})=e^{3p_0/\kappa}\label{param-conseq2}
\end{equation}
for any $t\in\mathcal{G}$, and further using the Weyl quantization, one easily finds \cite{PW2018, DS}  the expression for the starproduct defining the 4-d $\kappa$-Minkowski space together with the related involution. They are given by 
\begin{equation}
f\star g:=\mathcal{F}^{-1}(\mathcal{F}f\circ\mathcal{F}g),\ \ f^\dag:=\mathcal{F}^{-1}((\mathcal{F}f)^*)\label{basic-def1},
\end{equation}
for any $f,g\in\mathcal{F}(\mathcal{S}_c)$, with $f\star g\in\mathcal{F}(\mathcal{S}_c)$ and $f^\dag\in\mathcal{F}(\mathcal{S}_c)$. This gives rise to
\begin{align}
(f\star g)(x)&=\int \frac{dp^0}{2\pi}dy_0\ e^{-iy_0p^0}f(x_0+y_0,\vec{x})g(x_0,e^{-p^0/\kappa}\vec{x})  \label{starpro-4d},\\
f^\dag(x)&= \int \frac{dp^0}{2\pi}dy_0\ e^{-iy_0p^0}{\bar{f}}(x_0+y_0,e^{-p^0/\kappa}\vec{x})\label{invol-4d}.
\end{align}
 
To make contact with physical considerations, the eqns. \eqref{starpro-4d}, \eqref{invol-4d} must be extended to larger algebras which are subalgebras of the multiplier algebra of $\mathcal{M}^0_\kappa$. This has been nicely done in \cite{DS} in terms of an algebra of smooth functions with polynomial bounds together with all their derivatives which involves in particular the constants and the coordinate functions $x_\mu$. The corresponding algebra{\footnote{As vector space, it is a subspace of the vector space of the tempered distributions $\mathcal{S}^\prime(\mathbb{R}^4)$, as is $(\mathcal{F}(\mathcal{S}_c),\star)$.}}, denoted hereafter by $\mathcal{M}_\kappa^4$, is defined from the space of functions $f\in C^\infty(\mathbb{R}^4)$ satisfying the following two conditions{\footnote{We use the standard multi-index notation: $D^{m}_{\vec{x}}:=\prod_{j=1}^3\partial^{m_{j}}_j$, $m=(m_1,m_2,m_3)\in\mathbb{N}^3$}}:

\begin{equation}
f\in C^\infty(\mathbb{R}^4),\ \ \ |\partial^n_{x_0}D^m_{\vec{x}}f(x_0,\vec{x})|\le C_{n,m}(1+|x_0|)^{N_n}(1+|\vec{x}|)^{M_{n,m}},\label{polybounds}
\end{equation}
for any $n\in\mathbb{N}$, $m\in\mathbb{N}^3$, where $C_{n,m}>0$ and $N_n$, $M_{n,m}$ are numbers, and
\begin{equation}
\text{supp}(\mathcal{F}_{x_0}^{-1}(f))\ \ \text{compact},\label{compactsup}
\end{equation}
where $\mathcal{F}_{x_0}^{-1}$ denotes the Fourier transform with respect to the time variable and the symbol $\text{supp}$ its support. Let $\mathcal{B}(\mathbb{R}^4)$ denote the functions satisfying \eqref{polybounds} and \eqref{compactsup}. Hence, one has $\mathcal{M}_\kappa^4=(\mathcal{B}(\mathbb{R}^4),\star)$. \\
One can verify that $f\star g\in\mathcal{M}_\kappa^4$, $f^\dag\in\mathcal{M}_\kappa^4$ for any $f,g\in\mathcal{M}_\kappa^4 $ so that $\mathcal{M}_\kappa^4$ still defines a $^*$-algebra. By standard computations, one infers 
\begin{equation}\label{def-relations}
x_0\star x_i=x_0x_i+\frac{i}{\kappa}x_i,\ \ x_i\star x_0=x_0x_i,\ \ x_\mu^\dag=x_\mu,
\end{equation}
($i=1,2,3$, $\mu=0,\hdots,3$) consistent with the usual relations defining the $\kappa$-Minkowski space. \\ 

\subsection{$\kappa$-Poincar\'e invariance and twisted trace.}\label{kappatwist}

The right-invariant Haar measure related to the above convolution algebra provides a natural integration measure for building $\kappa$-Poincar\'e invariant action functionals on 
$\kappa$-Minkowski space. Indeed, upon using \eqref{param-conseq1} and further parametrizing the affine group elements in terms of momentum variables, the right-invariant Haar measure reduces simply to the Lebesgue measure \cite{PW2018, DS}
\begin{equation}
\mathcal{F}:d\nu(p_0,\vec{p})=d^4p\to d^4x\label{decadix}
\end{equation}
while for any action functional of the form $S_\kappa(\phi)=\int d^4x\ \mathcal{L}(\phi)$, one infers
\begin{equation}\label{invarquant}
h\blacktriangleright S_\kappa(\phi):=\int d^4x\ h\triangleright\mathcal{L}(\phi)=\epsilon(h)S_\kappa(\phi),
\end{equation}
for any $h$ in the $\kappa$-Poincar\'e Hopf algebra $\mathcal{P}_\kappa$, where $\epsilon: \mathcal{P}_\kappa\to\mathbb{C}$ is the counit of $\mathcal{P}_\kappa$ defined by $\epsilon(P_0)=\epsilon(P_i)=\epsilon(M_i)=\epsilon(N_i)=0,\  \epsilon(e^{-P_0/\kappa})=1 $, where $P_\mu$, $M_i$ and $N_i$ denote respectively the momenta, the rotations and the boosts. Relevant formulas for $\mathcal{P}_\kappa$ are given in the appendix \ref{apendixA}. Some useful formulas are
\begin{equation}\label{int-form1}
\int d^4x\ (f\star g^\dag)(x)=\int d^4x\ f(x){\bar{g}}(x)\,\mbox{ and } \,\int d^4x\ f^\dag(x)=\int d^4x\ {\bar{f}}(x).
\end{equation}
\\
It is known that the Lebesgue integral defines a twisted trace with respect to the star product \eqref{starpro-4d}. This can be equivalently rephrased by stating that the (positive) linear map{\footnote{A positive map verifies $\varphi(f)\ge0$ for $f\ge0$ .}} $\varphi:\mathcal{M}_\kappa^4\to\mathbb{C}$, $\varphi(f)=\int d^4x f(x)$, actually defines a KMS weight \cite{kuster} on $\mathcal{M}_\kappa^4$ for the group of $^*$-automorphisms of $\mathcal{M}_\kappa^4$ generated by $\sigma_t=e^{it\frac{3P_0}{\kappa}}=e^{t\frac{3\partial_0}{\kappa}}$, $t\in\mathbb{R}$ \cite{PW2018, DS, matas}.\\
Recall that a KMS weight on a ($C^*$-)algebra for a group of $^*$-automorphisms of the algebra, $\{\sigma_t\}_{t\in\mathbb{R}}$, called the modular (Tomita) group, is a positive linear map such that $\{\sigma_t\}_{t\in\mathbb{R}}$ has an analytic extension  $\{\sigma_z\}_{z\in\mathbb{C}}$ satisfying the two conditions{\footnote{for any $f$ in the domain of $\sigma_{\frac{i}{2}}$. Other more technical conditions must be verified by $\varphi$ and $\sigma_z$, namely $\varphi$ is lower semi continuous and $\{\sigma_z\}$ is norm-continuous \cite{kuster}.  Both are fulfilled in this paper.}}:
\begin{equation}
 i)\ \varphi\circ\sigma_z=\varphi,\ \ \ \  ii)\ \varphi(f^\dag\star f)=\varphi(\sigma_{\frac{i}{2}}(f)\star(\sigma_{\frac{i}{2}}(f))^\dag)\label{kmsweight}.
\end{equation}
For more mathematical properties of KMS weights, see \cite{kuster} and references therein{\footnote{In the sequel, it is understood that the notion of KMS weight is extended to the multiplier algebra of the C*-algebra of the affine group $C^*(\mathcal{G})$ (see e.g. the third ref. in \cite{kuster}). Since $\mathcal{G}$ is amenable, $C^*(\mathcal{G})$ is simply the completion of $L^1(\mathcal{G})$ w.r.t. the usual norm linked to the left regular representation on $L^2(\mathcal{G})$ and involves $\mathcal{M}^0_\kappa$ as a dense $^*$-subalgebra. Multipliers are then obtained by standard use of duality in the spaces of (temperated) distributions, whence the terminology KMS weight on $\mathcal{M}_\kappa$}}. \\
In this paper, we will mainly exploit the twisted trace aspects. First, the Lebesgue integral satisfies
\begin{equation}\label{twistrace}
\int d^4x\ (f\star g)(x)=\int d^4x\ \left((\sigma\triangleright g)\star f\right)(x),
\end{equation}
so that  the cyclicity with respect to the star product is lost. In \eqref{twistrace}, the twist $\sigma$ is given by
\begin{equation}\label{twistoperator}
\left(\sigma\triangleright f\right)(x_0,\vec{x}):=\left(e^{i\frac{3\partial_0}{\kappa}}\triangleright f\right)(x_0,\vec{x})=f(x_0+i\frac{3}{\kappa}, \vec{x}),
\end{equation}
or in terms of the generators of $\mathcal{P}_\kappa$ (see appendix \ref{apendixA})
\begin{equation}
\sigma=\mathcal{E}^3\label{twist-pratik}.
\end{equation}
In the following, we will call $\sigma$ \eqref{twist-pratik} the modular twist. \\
Now define $\sigma_z=e^{iz\frac{3P_0}{\kappa}}$, $z\in\mathbb{C}$. It is easy to verify that the set $\{\sigma_z\}_{z\in\mathbb{C}}$ forms a one-(complex) parameter group and that $\sigma_z(f\star g)=\sigma_z(f)\star\sigma_z(g)$ and moreover one has $\sigma_z(f^\dag)=(\sigma_{\bar{z}}(f))^\dag$ for any $f\in\mathcal{M}_\kappa^4$ and any $z\in\mathbb{C}$. Then, it is a simple matter of algebra to verify that the two conditions \eqref{kmsweight} for $\varphi(f)=\int d^4x f(x)$ to be a KMS weight on $\mathcal{M}_\kappa^4$ are verified. In view of $\sigma_z$, the relevant modular group $\{\sigma_t\}_{t\in\mathbb{R}}$ is generated by
\begin{equation}
\sigma_t=e^{it\frac{3P_0}{\kappa}}=e^{t\frac{3\partial_0}{\kappa}},\label{time-translat}
\end{equation}
for any $t\in\mathbb{R}$, which is the time-translation operator and defines $^*$-automorphisms since one has now $\sigma_t(f^\dag)=(\sigma_{{t}}(f))^\dag$ for any $f\in\mathcal{M}_\kappa^4$ and $t\in\mathbb{R}$. Eqn. \eqref{time-translat} can be expressed in terms of the Tomita operator $\Delta_T$ given by
\begin{equation}
\sigma_t=(\Delta_T)^{it},\, \mbox{ with } \,\Delta_T=e^{\frac{3P_0}{\kappa}} \label{tomita-op}.
\end{equation}
To make contact with the modular twist given in \eqref{twist-pratik}, simply notice that
\begin{equation}
\sigma=\sigma_{(t=i)}=(\Delta_T)^{-1}\label{lelien},
\end{equation}
which thus characterizes the modular twist occurring in the twisted trace formed by Lebesgue integral in terms of the generator of the modular automorphisms.
It can be easily verified that
\begin{equation}
\sigma(f\star g)=\sigma(f)\star\sigma(g),\label{aut-algebre}
\end{equation}
together with
\begin{equation}
\sigma(f^\dag)=(\sigma^{-1}(f))^\dag,\label{regulier-aut}
\end{equation}
the former relation \eqref{aut-algebre} showing that $\sigma$ is an automorphism of algebra, which however is a {\it{regular automorphism}} as signaled by \eqref{regulier-aut} but not a $^*$-automorphism. For mathematical details on regular automorphisms, see \cite{como-1}. Regular automorphisms appear naturally in the context of twisted spectral triples. For recent applications to noncommutative formulations of the 
Standard Model, see e.g. \cite{martinetti} and references therein.\\

In the subsequent analysis, we will need the expression for the star product and involution for d-dimensional $\kappa$-Minkowski spaces $\mathcal{M}_\kappa^d$. The corresponding extension of \eqref{starpro-4d} and \eqref{invol-4d} is straightforward and just amounts to replace in these equations spatial vectors $\vec{x}\in\mathbb{R}^3$ by spatial vectors $\vec{x}\in\mathbb{R}^{d-1}$, while the twist \eqref{twist-pratik} related to the twisted trace becomes 
\begin{equation}
\sigma_d=\mathcal{E}^{d-1}\label{twist-pratik-D},
\end{equation}
with obvious modifications in \eqref{twistoperator}, \eqref{tomita-op} and the integration measure \eqref{decadix}.

\section{Invariance and twisted trace.}\label{utomita}
In the subsection \ref{section22}, we collect the main properties of the framework describing the noncommutative connections on a right-module over an algebra. For a comprehensive review on noncommutative differential calculus and noncommutative extensions of the notion of connection, see \cite{mdv} and references therein. The details defining the unitary gauge group are briefly recalled in the appendix \ref{apend-gaugegroup}. For earlier works exploiting this framework for gauge theories on e.g. Moyal spaces $\mathbb{R}^4_\theta$ as well as $\mathbb{R}^3_\lambda$, see \cite{jcw-gauge1}-\cite{reviewgauge} and references therein. The extension of the present work to connections on bimodules \cite{mdv} is more involved and will be presented in a forthcoming publication. In the subsection \ref{section23}, we consider the case of {\it{untwisted}} noncommutative  differential calculus for the four dimensional $\kappa$-Minkowski space $\mathcal{M}_\kappa^4$. This encompasses most of the works on NCFT carried out so far. \\
Within this framework, we show that no polynomial real action depending on the curvature is invariant under the action of the usual gauge group\footnote{{Through misuse of language, we call ``gauge group'' the group of maps from the space-time to the gauge group.}} of maps from the space-time to $U\!\left(1\right)$, stemming from the occurrence of the twisted trace. However, such an action functional still remains invariant under a group of $U\!\left(1\right)$-valued maps invariant under the action of the Tomita operator.This result is then discussed.

\subsection{The gauge group.}\label{section22}

We introduce $\mathbb{E}$ a right-module over $\mathcal{M}_\kappa^4$ and $h_0$ a Hermitian structure chosen to be 
\begin{equation}
h_0(m_1,m_2)=m_1^\dag \star m_2,\label{ajout1}
\end{equation}
for any $m_1,m_2\in\mathbb{E}$ (for details, see appendix \ref{apend-gaugegroup}). In this paper, we will assume for convenience that $\mathbb{E}$ is one copy of the algebra $\mathcal{M}_\kappa^4${\footnote{Notice that we will continue below to denote the module by the symbol $\mathbb{E}$, despite it is a copy of the algebra $\mathcal{M}_\kappa^d$ in order to make the distinction between module and algebra apparent.}}, which turns $\mathbb{E}$ into a free module.\\
Recall that untwisted gauge transformations are usually defined as the set of automorphisms of $\mathbb{E}$ preserving its right-module structure over $\mathcal{M}_\kappa^4$ and compatible with the Hermitian structure $h_0$. From these requirements, one easily realizes that gauge transformations of any $\Phi\in\mathcal{M}_\kappa^4$ are simply given by 
\begin{equation}
\Phi^g=g\star \Phi\label{gaugematterbis}
\end{equation}
where $g\in\mathbb{E}$ verifies 
\begin{equation}
g^\dag\star g=g\star g^\dag=\bbone.\label{ajout2}
\end{equation}
The unitary gauge group is thus defined as
\begin{equation}
\mathcal{U}:=\{g\in\mathbb{E},\ \ g^\dag\star g=g\star g^\dag=\bbone  \}\label{unitar-groupbis},
\end{equation}
As far as a physical interpretation is concerned, first note that \eqref{gaugematterbis} can be viewed as gauge transformations of matter fields: the action of $\mathcal{U}$ on matter fields $\Phi$ in the algebra $\mathcal{M}_\kappa^4$ is simply the left multiplication of $\Phi$ by unitary elements of $\mathbb{E}\simeq\mathcal{M}_\kappa^4$. Besides, invariance of mass operators $\sim\Phi^\dag\star \Phi$ (where $\Phi$ is some scalar field) under the action of $\mathcal{U}$, namely $(\Phi^\dag\star\Phi)^g=\Phi^\dag\star\Phi$, is automatically achieved in view of the choice of the Hermitian structure \eqref{ajout1} together with \eqref{ajout2}.\\

We will look for $\kappa$-Poincar\'e invariant real actions, say $S_\kappa$, which depend {\it{polynomially}} on the curvature and satisfy the following properties:
\begin{enumerate}[a)]
\item $S_\kappa$ is invariant under the unitary gauge group $\mathcal{U}$ \eqref{unitar-groupbis} (or some related group),
\item $\lim_{\kappa\to\infty}S_\kappa$ coincides with standard Abelian gauge theory (in some d-dimensional Minkowski space).
\end{enumerate}
For the moment, we assume that the classical dimension of the $\kappa$-Minkowski space is equal to 4, i.e. we consider $\mathcal{M}_\kappa^4$. \\

Real action functionals are easily obtained, as it is the case for scalar NCFT \cite{PW2018}, by making use of the Hilbert product defined on $\mathcal{M}_\kappa^4${\footnote{The Hilbert space $\mathcal{H}$ is (unitarily) isomorphic to $L^2(\mathbb{R}^4)$ \cite{PW2018}; it is obtained by a standard GNS construction by completing (a subalgebra of) $\mathcal{M}_\kappa^4$ w.r.t. the canonical norm $||f||^2:=\langle f,f\rangle$. }} by
\begin{equation}
\langle f,g\rangle:=\int d^4x\left(f^\dag\star g\right)(x)=\int d^4x\ {\bar{f}}(x)(\mathcal{E}^3\triangleright g)(x),\  \forall f,g\in\mathcal{M}_\kappa^4.\label{hilbert-product}
\end{equation}
Observe that $\langle f,Kf\rangle\in\mathbb{R}$ for any self-adjoint operator $K$ and any $f\in\mathcal{M}_\kappa^4$. Indeed, one can write $\langle f,Kf\rangle={\overline{\langle Kf,f\rangle}}=\langle Kf,f\rangle$ where the first equality comes from the fact that $\langle.,.\rangle$ \eqref{hilbert-product} is a Hilbert product and the second one holds true since $K$ is assumed to be self-adjoint.\\
Therefore, owing to the fact that $\mathcal{P}_\kappa$ acts in a natural way on $\mathcal{M}_\kappa^4$, we will look for actions $S_\kappa$ of the form 
\begin{equation}
S_\kappa=\langle f,K\triangleright f\rangle \label{goal}
\end{equation}
with $K\in\mathcal{T}_\kappa$ (see appendix \ref{apendixA}) and $f$ related to the curvature, such that $\lim_{\kappa\to\infty}\langle f,K\triangleright f\rangle=S_{U(1)}$, $S_{U(1)}$ being the usual action for Abelian gauge theory.

\subsection{Group of invariance for actions with untwisted differential calculus.}\label{section23}

Let us assume that we have a noncommutative differential calculus defined (in obvious notations) by a graded differential algebra $(\Omega^\bullet,d)$ where $\Omega^\bullet=\oplus_{j\in\mathbb{N}}\Omega^j$ is a graded algebra{\footnote{The grading is as usual defined by the degree of forms. We assume that the involution \eqref{invol-4d} extends to $\Omega^\bullet$ and that $(d\omega)^\dag=d\omega^\dag$, which is satisfied by the differential calculi considered in this paper.}}, $\Omega^j$ corresponds to forms of degree $j$, $\Omega^0=\mathcal{M}_\kappa^4$, $d:\Omega^n\to\Omega^{n+1}$ is the (graded) differential fulfilling $d^2=0$ and the symbol $\times$ denotes the product of forms. The differential $d$ satisfies the Leibniz rule, assumed to be untwisted,
\begin{equation}
d(\omega\times\eta)=d\omega\times\eta+(-1)^{|\omega|}\omega \times d\eta\label{leibn-1}
\end{equation}
for any $\omega,\ \eta\in\Omega^\bullet$, where $|\omega|$ denotes the degree of $\omega$.\\

The action of the algebra $\mathcal{M}_\kappa^4$ on $\mathbb{E}\otimes\Omega^\bullet$, with $\mathbb{E}$ being a copy of $\mathcal{M}_\kappa^4$, is defined by $(m\otimes\omega).a=m\otimes(\omega\bullet a)$  for any $m\in\mathbb{E}$, $\omega\in\Omega^\bullet$, $a\in\mathcal{M}_\kappa^4$, in which
$\omega\bullet a$ denotes the extension of \eqref{module-action} defined as a right-multiplication by $a$ of the ``form components''. Accordingly, we set $\omega\bullet a=\omega\star a$. \\
Note by the way that the transformation rule with respect to a change of basis has to be understood now in the sense of the $\star$-product, that is, if $\omega = f\left(x\right) dx = g\left(y\right)dy$ and $y=h\left(x\right)$ then $f\left(x\right) = g\circ h\left(x\right)\star h'\left(x\right)$.\\

It can be realized that the above general features actually apply to the natural class of bicovariant differential calculi on $\kappa$-Minkowski, which are examples of practical relevance for application to Physics. To this class pertains in particular the $\kappa$-Poincar\'e invariant calculus which has been singled out in \cite{sit}. For a classification of the bicovariant differential calculi, see \cite{maj-diff}.\\

Recall that a Hermitian connection on a right-module \cite{mdv} can be defined as a map $\nabla:\mathbb{E}\to\mathbb{E}\otimes\Omega^1$ such that 
\begin{equation}
\nabla(ma)=\nabla(m)a+m\otimes da,\label{herm-conn-def} 
\end{equation}
for any $m\in\mathbb{E}$, $a\in\mathcal{M}_\kappa^4$ and
\begin{equation}
dh_0(m_1,m_2)=(\nabla(m_1),m_2)+(m_1,\nabla(m_2))\label{herm-conn-cond},
\end{equation}
with $h_0$ given here by \eqref{hermit-structure}. Since $\mathbb{E}\simeq\mathcal{M}_\kappa^4$ with $\mathcal{M}_\kappa^4$ unital, it is easy to realize that $\nabla$ is entirely determined by $\nabla(\bbone)$. Setting $\nabla(\bbone):=iA\in\Omega^1$, one has{\footnote{A factor $i$ is introduced to fit with most of the conventions of the physics literature.}}
\begin{equation}
\nabla(a)=iA\star a+da, \label{der-covar}
\end{equation}
with $A^\dag=A$ in view of \eqref{hermit-struc-cond} and $A\star a$ has been defined at the beginning of the subsection. The corresponding curvature is defined as a map 
\begin{equation}
F:\mathbb{E}\to\mathbb{E}\otimes\Omega^2,\ \ iF:=\nabla^2\label{curvature}
\end{equation}
leading to
\begin{equation}
F=dA+iA\times A\label{curvature-form}.
\end{equation}
More precisely, one could use the following equivalent definition: Consider the linear map $\hat{\nabla}:E \otimes \Omega^{1} \to E \otimes \Omega^{2}$ satisfying:
\begin{equation}
\hat{\nabla}\left(m\otimes\omega\right) = \nabla\left(m\right)\times\omega+m\otimes d\omega
\end{equation}
for any $m\in\mathbb{E}$, $\omega\in\Omega^1$. We define then the curvature operator to be the map $\hat{\nabla}\circ\nabla: E \to E \otimes \Omega^{2}$. For simplicity we drop the hat and write $\nabla^{2}$. The curvature $2$-form $F$ is then defined by $iF=\nabla^2\left(\bbone\right)$.\\

The unitary gauge transformations act in the usual way on the (affine) space of gauge connections, namely $\nabla^\phi=\phi^{-1}\circ\nabla\circ\phi$
for any $\phi\in\text{Aut}_{h_0}(\mathbb{E})$ (see appendix \ref{apend-gaugegroup}), from which one easily finds
\begin{equation}
\nabla^g(a)=iA^g\star a+da,\ \ A^g=g^\dag \star A \star g+g^\dag \star dg,\label{gaugetrans-conn},
\end{equation}
together with
\begin{equation}
F^g=g^\dag \star F\star g\label{gaugetrans-curv},
\end{equation}
for any $g\in\mathcal{U}$ and any $a\in\mathcal{M}_\kappa^4$. Note that $\mathcal{U}$ acts on $\mathbb{E}\otimes\Omega^\bullet$ as $g\otimes\bbone_{\Omega^\bullet}$.\\

We now look for gauge invariant real actions with polynomial dependence in the curvature. \\

Consider first the representative case of a $\kappa$-Poincar\'e invariant differential calculus \cite{sit}. Let $(e^\mu)_{(\mu=0,\hdots,4)}$ be the 5-dimensional base of the forms whose components, however, are still functions of $\mathcal{M}_\kappa^4$ (hence depending on 4 variables).\\

Within this differential calculus, one can define a natural notion of integration of forms{\footnote{The twisted trace can be used to construct a twisted K-cycle over the algebra leading to a noncommutative analog of integration and noncommutative analogue of Hodge duality on forms \cite{mercat-sit}.}}. In particular, for any 5-form $\omega_5$, one has $\int \omega_5=\int d\mu_5[\omega_5]=\int d^4x\ [\omega_5](x)$ where $d\mu_5$ is a volume form and $[\omega_5](x)\in\mathcal{M}_\kappa^4$ is the ``component'' of the 5-form $\omega_5$ \cite{mercat-sit}. Furthermore, $\Omega^\bullet$ can be equipped with an inner product 
\begin{equation}
\langle\omega, \eta\rangle=\int \omega^\dag\times(\tilde{\star}\eta)=\int d^4x\ [\omega^\dag\times(\tilde{\star}\eta)](x),\label{innerprod-forms}
\end{equation}
which thus extends \eqref{hilbert-product}. Here, the symbol $\tilde{\star}$ denotes a noncommutative analogue of the Hodge operation: $\tilde{\star}:\Omega^n\to\Omega^{5-n}$, $n=0,...,5$, not to be confused with the star product. It is defined \cite{mercat-sit}, in obvious notations, by:
\begin{equation}
\tilde{\star}(\bbone)=d\mu_5,\ \tilde{\star}(e^\mu)=\frac{1}{4!}\delta^{\mu\nu}\epsilon_{\nu\rho\lambda\theta\zeta}e^\rho\times e^\lambda\times e^\theta\times e^\zeta,\ \tilde{\star}(e^\mu\times e^\nu)=\frac{1}{3!}\delta^{\mu\gamma}\delta^{\nu\delta}\epsilon_{\gamma\delta\lambda\zeta\theta}e^\lambda\times e^\zeta\times e^\theta\label{hodge}.
\end{equation}
In view of \eqref{gaugetrans-curv} and the discussion given at the end of the subsection \ref{section22}, a natural candidate for a real action of a gauge theory is then
\begin{equation}
S_\kappa=\langle F,F\rangle=\int d^4x\ F_{\mu\nu}^\dag\star F_{\mu\nu}\label{ncyang-1}
\end{equation}
where \eqref{innerprod-forms} and \eqref{hodge} have been used to obtain the second equality and $F_{\mu\nu}$ is the component of the 2-form curvature, namely $F=\frac{1}{2}F_{\mu\nu}e^\mu\times e^\nu$. $S_\kappa$ is obviously $\kappa$-Poincar\'e invariant, stemming from \eqref{invarquant}.  It can be easily verified that one has formally $\lim_{\kappa\to\infty}S_\kappa=S_{U(1)}$.\\
Now, from \eqref{gaugetrans-curv}, one infers that
\begin{equation}
\langle F^g,F^g\rangle=\int d^4x\  (\mathcal{E}^3(g)\star g^\dag)\star F_{\mu\nu}^\dag \star F_{\mu\nu}\label{transfo-1},
\end{equation}
for any $g\in\mathcal{U}$, so that $\langle F^g,F^g\rangle$ is invariant under any unitary gauge transformation $g\in\mathcal{U}$ satisfying the condition
\begin{equation}
\mathcal{E}^3(g)\star g^\dag=1\label{condition-invar},
\end{equation}
which, owing to the fact that $g^\dag\star g=\bbone$, can be (equivalently) rewritten as
\begin{equation}
\mathcal{E}^3(g)=g\label{cond-invarKMS}.
\end{equation}
By using \eqref{tomita-op} and \eqref{twistoperator}, eqn. \eqref{cond-invarKMS} can be expressed in terms of the Tomita operator $\Delta_T$ \eqref{tomita-op} as
\begin{equation}
\Delta_T(g)=g\label{tomita-invar}.
\end{equation}
Let us summarize the above discussion. Within the above $\kappa$-Poincar\'e invariant calculus, we find that the most natural noncommutative extension of the (commutative) Abelian gauge theory provided by the polynomial action \eqref{ncyang-1} is not invariant under the unitary gauge group $\mathcal{U}$ \eqref{unitar-groupbis}. \\
However, it is still invariant under the group $\mathcal{U}_\Delta$ involving, in particular, all the elements $g(x_0;\vec{x})$ of $\mathcal{U}$, i.e. the unitary elements of $\mathcal{M}_\kappa^4$, admitting a continuation on some domain $U\subseteq\mathbb{C}$, $\tilde{g}(z;\vec{x})$ with $\mathfrak{R}(z)=x_0$, such that according to \eqref{tomita-invar} $\Delta_T(\tilde{g}(\mathfrak{R}(z);\vec{x}))=\Delta_T(g(x_0;\vec{x}))=g(x_0;\vec{x})$. Having this in mind, we simply set
\begin{equation}
\mathcal{U}_\Delta:=\{g\in\mathcal{U},\  \Delta_T(g)=g\}\label{invariance-group}.
\end{equation}
At this point, some comments are in order:
\begin{enumerate}
\item If one is willing to consider seriously the (gauge) invariance related to $\mathcal{U}_\Delta$, it appears that the fulfillment of \eqref{tomita-invar} is a very strong constraint on the allowed gauge transformations which may severely restrict the content of $\mathcal{U}_\Delta$ unless the algebra $\mathcal{M}_\kappa^4$ is suitably enlarged, a task which is beyond the scope of this paper.\\
Note that any element $g$ depending only on the {\it{spatial}} coordinates belongs to $\mathcal{U}_\Delta$. This immediately follows from \eqref{twistoperator}. Upon using \eqref{starpro-4d} and \eqref{invol-4d} for time-independent functions $g(\vec{x})$, one easily realizes that any element $g(\vec{x})\in\mathcal{U}_\Delta$ is of the form $g(\vec{x})=e^{i\omega(\vec{x})}$ with $\omega(\vec{x})\in\mathcal{M}_\kappa^4$.\\
In the same way, one easily realizes that gauge transformations depending only on time $g(x_0)\in\mathcal{U}_\Delta$ must be of the form 
\begin{equation}
g(x_0)=e^{i\varphi({x_0})},\label{periodic-im1}
\end{equation}
while, denoting $\tilde{g}(z)=e^{i\tilde{\varphi}(z)}$ with $\mathfrak{R}(z)=x_0$, a continuation of $g$ on some domain $U\subseteq\mathbb{C}$, one must have
\begin{equation}
 \mathcal{E}^3\triangleright\tilde{\varphi}({x_0})=\tilde{\varphi}({x_0+i\frac{3}{\kappa}})=\tilde{\varphi}({x_0})\label{periodic-im2}
\end{equation}
which however cannot necessarily fulfill the condition \eqref{compactsup} (unless $\varphi(x_0)$ is constant). For instance, pick $g(x_0)=e^{i\tanh(x_0)}$; it verifies \eqref{periodic-im1} and \eqref{periodic-im2} (as well as \eqref{polybounds}) but the Fourier transform is not compactly supported so that \eqref{compactsup} is not satisfied. Hence $g(x_0)$ \eqref{periodic-im1} does not belong to $\mathcal{M}_\kappa^4$ unless the constraints \eqref{polybounds}, \eqref{compactsup} are revisited.\\
Examining how $\mathcal{M}_\kappa^4$ can actually be enlarged in such a way that it allows for the existence of non trivial time-depending gauge transformations satisfying \eqref{tomita-invar} is beyond the scope of this paper. In the section \ref{section3}, we will follow an alternative more algebraic route which gives rise to polynomial functionals on $\mathcal{M}_\kappa^d$ invariant under the full gauge group $\mathcal{U}$. 

\item Notice that invariance under the full gauge group $\mathcal{U}$ of functionals $\sim\langle f,K\triangleright f\rangle$ where $K\in\mathcal{P}_\kappa$ (with $K$ a morphism of algebra and $[K,\mathcal{E}]=0$, the case of practical interest) cannot be achieved. Indeed, one has
\begin{equation}
\int d^4x\ F_{\mu\nu}^{g^\dag}\star K\triangleright F_{\mu\nu}^g=\int d^4x\ \mathcal{E}^3K(g)\star g^\dag\star F_{\mu\nu}^{\dag}\star g\star K\triangleright g^\dag\star  F_{\mu\nu}
\end{equation}
so that the invariance is realized provided $ \mathcal{E}^3K\triangleright g\star g^\dag=1$ and $g\star K\triangleright g^\dag=1$. It follows that $K\triangleright(\mathcal{E}^3\star g^\dag)=1$ must hold true which however is not verified for any $g\in\mathcal{U}$.
\item The above conclusion also applies to any differential calculus whose differential is untwisted, as described at the beginning of this subsection. Such a framework leads to untwisted gauge transformations, as given by \eqref{gaugetrans-conn}, \eqref{gaugetrans-curv}. This holds true in particular for the natural class of bicovariant differential calculi on $\kappa$-Minkowski. Requiring the $\kappa$-Poincar\'e invariance forces the use of the Lebesgue integral behaving as a twisted trace in the polynomial action functional{\footnote{This may come either from a proper extension of the inner product \eqref{innerprod-forms} or simply ``put by hand'' for practical purpose in the action functional.}}, by eqn. \eqref{invarquant}. This leads, upon gauge transformation, to a behaviour similar to the one described by \eqref{transfo-1}-\eqref{tomita-invar}, which arises simply because there is nothing in the present untwisted framework that may compensate the contribution stemming from the modular twist $\mathcal{E}^3(g)$ showing up when unitary gauge factors $g,\ g^\dag$ are permuted, due to \eqref{twistrace}.

\item Through this paper, we use noncommutative connections on a right-module which is the most often used description in the NCFT literature. The case of noncommutative (linear) connections on a bimodule \cite{mdv} is more involved and needs further investigation to be carried out. The corresponding analysis has been undertaken and will be presented in a forthcoming publication.

\end{enumerate}

We now consider another more algebraic route which will lead to polynomial functionals invariant under the full gauge group $\mathcal{U}$. This can be achieved by selecting a suitable twisted differential calculus and extending the notion of noncommutative connection in such a way that the net effects on the resulting (hence twisted) gauge transformations compensate the effect of the modular twist $\mathcal{E}^3=\Delta_T^{-1}$ related to the integral \eqref{twistrace}. \\
Recall that twisted gauge transformations show up whenever e.g. the differential $d$ is twisted. Namely one has now the following twisted Leibniz rule (in obvious notations) $d(\omega\times \eta)=d\omega\times \eta+(-1)^{|\omega|}\rho(\omega)\times d\eta$ for any $\omega,\ \eta\in\Omega^\bullet$, in which $\rho:\mathcal{M}_\kappa^d\to\mathcal{M}_\kappa^d$ is some (auto)morphism of $\mathcal{M}_\kappa^d$. This generally forces the occurence of a twist in the gauge transformations in order to ensure the stability of the space of connections under the gauge group action. We note that twisted structures in noncommutative geometry appear in the context of Twisted Spectral Triples \cite{como-1}. Twisted Spectral Triples also appeared very recently within the context of the noncommutative formulation of the Standard Model (see e.g. \cite{martinetti} and references therein). \\

\section{Gauge-invariant models from twisted connections.}\label{section3}

In this section, we will select a natural Abelian Lie algebra of (bi)twisted derivations belonging to the so-called ``deformed translation algebra'' $\mathcal{T}_\kappa\subset\mathcal{P}_\kappa$. These twisted derivations are sometimes called ($\tau$,$\sigma)$-derivations in the mathematical literature where $\tau$ and $\sigma$ are morphisms deforming the usual Leibniz rule characterizing the derivations. They appeared in particular in the context of Ore extensions \cite{ore} and Hom-Lie algebras \cite{Hom-Lie}.\\

We move from the 4- to the d-dimensional $\kappa$-Minkowski space $\mathcal{M}_\kappa^d$ and consider a set of $d$ twisted derivations. The algebra for $\mathcal{M}_\kappa^d$ is defined by the obvious replacement $\mathbb{R}^4\rightarrow\mathbb{R}^d$ in \eqref{polybounds}-\eqref{def-relations} and below, while the modular twist is given by \eqref{twist-pratik-D}. The expressions for the corresponding star product and involution are defined in the last paragraph of the subsection \ref{kappatwist}. \\
To clarify the notations and avoid confusion between twists \eqref{twistrace}, in particular the modular twist, we now change the usual notations of the literature and replace from now on the symbols $\tau$ and $\sigma$ respectively by $\alpha$ and $\beta$. We will call twisted derivations (resp. bitwisted) the $(\bbone,\beta)$-derivations (resp. the $(\alpha,\beta)$-derivations).\\

In the subsection  \ref{section31}, we generalize the by-now standard notion of derivation-based differential calculus \cite{mdv} and construct a differential calculus based on bitwisted derivations which formally reduces to the usual de Rham differential calculus at the commutative limit $\kappa\to\infty$. Notice that this can be viewed as a further extension of \cite{jcw-gauge2} where the so-called $\epsilon$-derivations were considered. \\
We then consider the case of twisted derivations-based differential calculus in the subsection \ref{section32} and extend the notion of noncommutative connection on a (right-)module to twisted connections together with their gauge transformations. This generalizes the notion of $\varepsilon$-connection developed in \cite{jcw-gauge2}.\\
Then looking for action functionals of the form \eqref{goal} invariant under the full gauge group \eqref{unitar-groupbis}, we show in the subsections \ref{section43} and \ref{section44} the existence of a strong relation between gauge-invariance of the action, the classical dimension of the $\kappa$-Minkowski space and the properties of the twists of the chosen differential calculus. In particular, we show that within the framework developed in the subsection \ref{section43}, the coexistence of the $\kappa$-Poincar\'e invariance and the gauge invariance implies that the unique value for the classical dimension of the $\kappa$-Minkowski space is equal to 5. These results are then discussed.

\subsection{Noncommutative differential calculus based on twisted derivations. }\label{section31}

In the following, $\alpha$ and $\beta$ will be assumed to be regular automorphisms of $\mathcal{M}_\kappa^d$. Recall that a bitwisted ($\alpha$,$\beta)$-derivation $X$ of $\mathcal{M}_\kappa^d$ is defined as a map $X:\mathcal{M}_\kappa^d\to\mathcal{M}_\kappa^d$ satisfying 
\begin{equation}
X(a\star b)=X(a)\star \alpha(b)+\beta(a)\star X(b),\label{Leibniz-sigtau}
\end{equation}
for any $a,b\in\mathcal{M}_\kappa^d$, that is, a derivation whose Leibniz rule is twisted by the two automorphisms $\alpha$ and $\beta$ of $\mathcal{M}_\kappa^d$.\\
Assume that one has a family of such bitwisted derivations, $X_\mu,\ \mu=0,...,(D-1)$ generating an Abelian Lie algebra, i.e. $[X_\mu,X_\nu]=0$ denoted by $\mathfrak{D}$. Here, $D$ is not necessarely equal to $d$.\\
It turns out that the notion of noncommutative differential calculus based on a Lie algebra of derivations of an associative algebra can be straightforwardly extended to a differential calculus based on an Abelian Lie algebra of bitwisted derivations, together with the notion of connection on a (right-)module presented in the next subsections.\\

 Let $\Omega^n(\mathfrak{D},\mathbb{E})$, $n\in\mathbb{N}$, be the linear space of $n$-$Z(\mathcal{M }_\kappa^d)$-linear antisymmetric forms with value in $\mathbb{E}$, $\omega:\mathfrak{D}^n\to\mathbb{E}$ satisfying 
\begin{equation}
\omega(X_1,X_2,...,X_n)\in\mathbb{E},\label{formule1}
\end{equation}
\begin{equation}
\omega(X_1,X_2,...,X_n.z)=\omega(X_1,X_2,...,X_n)\star z\label{formule2},
\end{equation}
for any $z$ in $Z(\mathcal{M }_\kappa^d)$, the center of $\mathcal{M }_\kappa^d$ and any $X_1,...X_n\in\mathfrak{D}$. In \eqref{formule2}, $X.z$ is defined by $(X.z)(a):=X(a)\star z=z\star X(a)=(z.X)(a)$ for any $a\in\mathcal{M}_\kappa^d$.  Note that $\mathfrak{D}$ carries a bimodule structure over $Z(\mathcal{M }_\kappa^d)$.\\

 We set as usual
\begin{equation}
\Omega^\bullet:=\bigoplus_{n=0}^{D-1}\Omega^n(\mathfrak{D},\mathbb{E}), 
\end{equation}
 and $\Omega^0=\mathcal{M}_\kappa^d$.\\

Then, it is a simple matter of standard computation to verify that the set of data $(\Omega^\bullet,{\bf{d}})$ defines a differential algebra with the product $\times$ and the differential $\bf{d}$ satisfying, as usual:
\begin{equation}
{\bf{d}}:\Omega^p(\mathfrak{D},\mathbb{E})\to\Omega^{p+1}(\mathfrak{D},\mathbb{E}),\ \forall p\in\{0,...,(D-1) \}
\end{equation}
\begin{equation}
{\bf{d}}^2=0, \label{nilpot}
\end{equation}
are respectively given by
\begin{align}
\nonumber
&(\omega\times\eta)(X_1,...,X_{p+q})\\
&\qquad\qquad=\frac{1}{p!q!}\sum_{s\in\mathfrak{S}(p+q)}(-1)^{\text{sign}(s)}\omega(X_{s(1)},...,X_{s(p)})\star \eta(X_{s(p+1)},...,X_{s(q)})\label{ncwedge1},
\end{align}
and
\begin{equation}
{\bf{d}}\omega(X_1,X_2,...,X_{p+1})=\sum_{i=1}^{p+1}(-1)^{i+1}X_i\big(\omega(X_1,...,\vee_i,...,X_{p+1})\big), \label{ncwedge2}
\end{equation}
for any $\omega\in\Omega^p(\mathfrak{D},\mathbb{E})$, $\eta\in\Omega^q(\mathfrak{D},\mathbb{E})$ and any elements of $\mathfrak{D}$, where in \eqref{ncwedge1} $\mathfrak{S}(p+q)$ denotes as usual the symmetric group of $p+q$ elements, $\text{sign}(s)$ is the signature of the permutation $s$ and in \eqref{ncwedge2} the symbol $\vee_i$ denotes the omission of the element $X_i$.\\

Keeping in mind the duality between $\mathcal{T}_\kappa$ and $\mathcal{M}_\kappa^d$ and the fact that one looks for actions with suitable commutative limit, it is natural to focus on those $X_\mu$ verifying $X_\mu\in\mathcal{T}_\kappa$. This implies a strong restriction on the possible allowed form for these objects.\\
Indeed, if $X_\mu\in\mathcal{T}_\kappa$ verifies \eqref{Leibniz-sigtau}, then one must also have 
\begin{equation}
\Delta(X_\mu)=X_\mu\otimes\alpha+\beta\otimes X_\mu\label{deltaX}
\end{equation}
with $\alpha\in\mathcal{T}_\kappa$ and $\beta\in\mathcal{T}_\kappa$ so that $[X_\mu,\alpha]=0$, $[X_\mu,\beta]=0$ since $\mathcal{T}_\kappa$ is Abelian. \\
Since $\alpha$ and $\beta$ must be compatible with the structures of algebra in $\mathcal{M}_\kappa^d$, one must have $\Delta(\alpha)=\alpha\otimes\alpha$ and $\Delta(\beta)=\beta\otimes\beta$. This, combined to the fact that $\mathcal{T}_\kappa$ is generated by $\mathcal{E}$ (together with the $P_\mu$'s) implies $\alpha=\mathcal{E}^{x_1}$, $\beta=\mathcal{E}^{x_2}$ where $x_1$ and $x_2$ are two real numbers. Since any element of $\mathcal{T}_\kappa$ is a finite linear combination of powers of the generators and that any $X_\mu$ must obey a twisted Leibniz rule as \eqref{Leibniz-sigtau}, the possible allowed combinations are of the form $X_0=\sum a_{u_1v_1}\mathcal{E}^{u_1}P_0^{v_1}$ and $X_i=\sum b_{u_2v_2}\mathcal{E}^{u_2}P_i^{v_2}$ where $ a_{u_1v_1},\  b_{u_2v_2}$ are constants and the $u,v$'s are numbers. Finally, plugging these 2 expressions into \eqref{deltaX} and identifying the left- and right-handside singles out the unique following one-parameter family of  bitwisted derivations:
\begin{equation}
X_0=\kappa\mathcal{E}^\gamma(1-\mathcal{E})\,\mbox{ and }\,X_i=\mathcal{E}^\gamma P_i\,\mbox{ for }\, i=1,...,(d-1)\label{sigtau-famil}
\end{equation}
where $\mathcal{E},P_i\in\mathcal{P}_\kappa$ (see appendix \ref{apendixA}) and $\gamma$ is a real parameter to be fixed in a while). \\

We denote $\mathfrak{D}_\gamma$ the Abelian Lie algebra of twisted derivations generated by \eqref{sigtau-famil}. It can be easily verified by using the relations \eqref{deriv-twist1}-\eqref{relation-calE} that $\alpha$ and $\beta$ do not depend on $\mu$ and that the $X_\mu$'s \eqref{sigtau-famil} verify $\alpha= \mathcal{E}^\gamma$ and $\beta=\mathcal{E}^{\gamma+1}$, stemming from
\begin{equation}
X_\mu(a\star b)=X_\mu(a)\star (\mathcal{E}^\gamma\triangleright b)+ (\mathcal{E}^{1+\gamma}\triangleright a)\star X_\mu(b).\label{tausigleibgene}
\end{equation}
The operators $X_\mu$, $\mu=0,..,(d-1)$ are self-adjoint. For instance, standard algebraic manipulations using \eqref{hopf1}-\eqref{hopf4bis}, \eqref{deriv-twist1} and \eqref{invarquant} yield
\begin{eqnarray}
\langle X_i(f),g\rangle&=&\int d^dx\ S(\mathcal{E}^\alpha P_i)^\dag\triangleright f^\dag\star g=-\int d^dx\ \mathcal{E}^{-1-\alpha}P_i\triangleright f^\dag\star g\nonumber\\
&=&-\int d^dx\ P_i\triangleright f^\dag\star \mathcal{E}^{1+\alpha}g=\int d^dx\ f^\dag\star  \mathcal{E}^{\alpha}P_i\triangleright g
=\langle f,X_i(g)\rangle.
\end{eqnarray}
 for any $f,g\in\mathcal{M}_\kappa^d$. One proceeds similarly for $X_0$.\\

We observe that the $X_\mu$'s \eqref{sigtau-famil} are closely related to the operators used to build the (2-d and 4-d) Dirac operator involved in the modular spectral triple presented in \cite{matas}. In particular, the case $\gamma=0$ exhibits a salient feature: the corresponding twisted derivations are the unique elements of $\mathcal{T}_\kappa$ linked to a Dirac operator with suitable properties for a modular spectral triple \cite{matas}. Notice that the mass dimension of the $X_\mu$'s is $[X_\mu]=1$. Moreover, one obtains, using \eqref{poinc1}-\eqref{poinc3}
\begin{equation}
[X_\mu,X_\nu]=0\,\mbox{ and }\,\lim_{\kappa\to0}X_\mu=P_\mu\,\mbox{ for }\,\mu,\nu=0,...,(d-1)\label{lim-translat},
\end{equation}
so that the Abelian Lie algebra generated by \eqref{sigtau-famil} coincides with the usual Lie algebra of translations in the commutative limit.\\

\subsection{Twisted connections.}\label{section32}

In this subsection, we assume that we have $\gamma=0$. Hence, $\alpha=\bbone$, $\beta=\mathcal{E}$ and the twisted derivations \eqref{sigtau-famil} generating $\mathfrak{D}_0$ reduce to 
\begin{equation}
X_0=\kappa(1-\mathcal{E}),\ \ X_i=P_i,\ \ i=1,...,(d-1)\label{modularDirac}
\end{equation}
with  the twisted Leibniz rule \eqref{tausigleibgene} simplifying into $X_\mu(a\star b)=X_\mu(a)\star b+ (\mathcal{E}\triangleright a)\star X_\mu(b)$.\\
The underlying differential calculus is assumed to be of the type presented in the subsection \ref{section31} where the Abelian Lie algebra of twisted derivations is the one generated by \eqref{modularDirac}. Note that one must have now $D=d$.\\

Given $\mathfrak{D}_0$, we define a twisted connection as a map such that for any $X_\mu\in\mathfrak{D}_0$, $\nabla_{X_\mu}:\mathbb{E}\to\mathbb{E}$ ($\mathbb{E}$ is still a copy of $\mathcal{M}_\kappa^d$) satisfies:
\begin{eqnarray}
\nabla_{X_\mu+X^\prime_\mu}(m)&=&\nabla_{X_\mu}(m)+\nabla_{X^\prime_\mu}(m)\label{sigtaucon1}\\
\nabla_{z.X_\mu}(m)&=&\nabla_{X_\mu}(m)\star z\label{sigtaucon2}\\
\nabla_{X_\mu}(m\star a)&=&\nabla_{X_\mu}(m)\star a+\tilde{\beta}(m)\star X_\mu(a)\label{sigtaucon3},
\end{eqnarray}
for any $m\in\mathbb{E}$, $X_\mu,X^\prime_\mu\in\mathfrak{D}_0$, $z\in Z(\mathcal{M }_\kappa^d)$, $a\in\mathcal{M}_\kappa^d$. In \eqref{sigtaucon3}, we have introduced a morphism $\tilde{\beta}:\mathbb{E}\to\mathbb{E}$ whose action on the module is simply defined by $\tilde{\beta}(m)=\beta(m)=\mathcal{E}\triangleright m$ for any $m$ in $\mathbb{E}\simeq\mathcal{M}_\kappa^d$.\\

From \eqref{sigtaucon3}, it follows that
\begin{equation}
\nabla_{X_\mu}(a)=A_{X_\mu}\star a+X_{\mu}(a),\label{derivcov}
\end{equation}
\begin{equation}
A_{X_\mu}:=\nabla_{X_\mu}(\bbone)\label{gaugepot}.
\end{equation}
Now, for any ``matter'' gauge transformation \eqref{gaugematterbis}, one can write
\begin{eqnarray}
\nabla_{X_\mu}(g\star a)&=&\nabla_{X_\mu}(g)\star a+(\mathcal{E}\triangleright g)\star X_\mu(a)\nonumber\\
&=&(A_{X_\mu}\star g+X_\mu(g))\star a+(\mathcal{E}\triangleright g)\star X_\mu(a)\label{intermedd}.
\end{eqnarray}
Thus, we define the gauge transformations of the twisted connection as 
\begin{equation}
\nabla^\phi:=(\beta\circ\phi^{-1})\circ\nabla\circ\phi
\end{equation}
for any $\phi\in\text{Aut}_{h_0}$ (see appendix \ref{apend-gaugegroup}) or equivalently
\begin{equation}
\nabla^g_{X_\mu}(a):=(\mathcal{E}\triangleright g^\dag)\star\nabla_{X_\mu}(g\star a),\ \ \forall g\in\mathcal{U},\ \forall a\in\mathcal{M}_\kappa^d\label{twistedgauge},
\end{equation}
which therefore represents a twisted gauge transformation related to the twist $\beta=\mathcal{E}$.\\

It simply follows that
\begin{equation}
\nabla^g_{X_\mu}(a)=A^g_{X_\mu}\star a+X_{\mu}(a)
\end{equation}
with
\begin{equation}
A^g_{X_\mu}=(\mathcal{E}\triangleright g^\dag)\star A_{X_\mu}\star g+(\mathcal{E}\triangleright g^\dag)\star X_\mu(g)\label{transfgaugpot},
\end{equation}
for any $g\in\mathcal{U}$, which thus differs from the untwisted gauge transformations of the noncommutative gauge potential by the presence of the twist $\beta=\mathcal{E}$.
We set
\begin{equation}
A_\mu:=A_{X_\mu}.
\end{equation}
Using the framework defining the general twisted differential calculus of subsection \ref{section31}, the above expressions easily extend to $\nabla:\mathbb{E}\to\mathbb{E}\otimes\Omega^1$ (with $\Omega^j:=\Omega^j(\mathfrak{D}_0,\mathbb{E})$, for any $j\in\{0,...,(d-1)\}$) and one obtains in particular
\begin{equation}
\nabla(a)=A\star a+ 1\otimes da,\ \ A^g=(\mathcal{E}\triangleright g^\dag)\star A \star g+(\mathcal{E}\triangleright g^\dag)\star dg,\label{forms-alph1}
\end{equation}
with $A\in\Omega^1$, $d$ can be straightforwardly defined from \eqref{ncwedge2} with $d^2=0$ thanks to the fact that the $X_\mu$'s commute with each other. In \eqref{forms-alph1} we have used the explicit expression for the action of the algebra on the module (see below eqn. \eqref{leibn-1}).\\

Now, the curvature related to the twisted connection is defined as a map such that for any $X_\mu,X_\nu\in\mathfrak{D}_0$, $F(X_\mu,X_\nu):\mathbb{E}\to\mathbb{E}$ with
\begin{equation}
F_{\mu\nu}:=F(X_\mu,X_\nu)=\nabla_{X_\mu}\circ\beta^{-1}\circ \nabla_{X_\nu}-\nabla_{X_\nu}\circ\beta^{-1}\circ \nabla_{X_\mu}\label{curvature1},
\end{equation}
with still $\beta=\mathcal{E}$. One has $F_{\mu\nu}(m\star a)=F_{\mu\nu}(m)\star \beta^{-1}(a)$. We set
\begin{equation}
F_{\mu\nu}=X_\mu(\mathcal{E}^{-1}\triangleright A_\nu)-X_\nu(\mathcal{E}^{-1}\triangleright A_\mu)+A_\mu\star(\mathcal{E}^{-1}\triangleright A_\nu)-A_\nu\star(\mathcal{E}^{-1}\triangleright A_\mu)\label{fmunu1}.
\end{equation}
Notice that this latter expression extends to $\Omega^2$, namely from the subsection \ref{section31}, one has
\begin{equation}
F=d\beta^{-1}(A)+A\times\beta^{-1}(A),
\end{equation}
where $(\beta^{-1}\omega)(X_1,...,X_p)=\beta^{-1}(\omega(X_1,...,X_p))$ for any $\omega\in\Omega^p$ and still $\beta=\mathcal{E}$.\\

Now, combining \eqref{fmunu1} with \eqref{transfgaugpot}, a standard computation yields
\begin{equation}
F_{\mu\nu}^g=\beta(g^\dag)\star F_{\mu\nu}\star \beta^{-1}(g)=(\mathcal{E}\triangleright g^\dag)\star 
F_{\mu\nu}\star (\mathcal{E}^{-1}\triangleright g)\label{fg1},
\end{equation}
for any $g\in\mathcal{U}$.\\

The extension of the present construction to any Abelian Lie algebra $\mathfrak{D}$ of twisted derivations \eqref{Leibniz-sigtau} with $\alpha=\bbone$ is straightforward.

\subsection{Fixing space-time dimension from gauge invariance requirement.}\label{section43}
Given the Abelian Lie algebra of twisted derivations $\mathfrak{D}_0$, together with the corresponding differential calculus and related twisted connection and curvature derived in the subsection \ref{section32}, we now look for gauge-invariant action functionals of the form
\begin{equation}
S_\kappa=\int d^dx \ F^\dag_{\mu\nu}\star J(F_{\mu\nu})\label{action1},
\end{equation}
where $J\in\mathcal{T}_\kappa$ with $J(a\star b)=J(a)\star J(b)$, to be determined. Thus, by performing a gauge transformation \eqref{fg1} on \eqref{action1}, one can write
\begin{eqnarray}
S_\kappa^g&=&\int d^dx\ ( F^g)^\dag_{\mu\nu}\star J( F^g_{\mu\nu})\nonumber\\
&=\!&\!\int d^dx\ (\mathcal{E}\triangleright g^\dag)\star  F^\dag_{\mu\nu}\star(\mathcal{E}^{-1}\triangleright g) \star J(\mathcal{E}\triangleright g^\dag)\star J(F_{\mu\nu})\star J(\mathcal{E}^{-1}\triangleright g).
\end{eqnarray}
where we used the fact that all elements of $\mathcal{T}_\kappa$ commute with each other. By further making use of \eqref{twistrace} and the expression for the modular twist in d-dimension \eqref{twist-pratik-D}, one obtains
\begin{equation}
S_\kappa^g=\int d^dx\ J(\mathcal{E}^{d-2}\triangleright g)\star (\mathcal{E}\triangleright g^\dag)\star  F^\dag_{\mu\nu}\star(\mathcal{E}^{-1}\triangleright g) \star J(\mathcal{E}\triangleright g^\dag)\star J(F_{\mu\nu}).
\end{equation}
Then, gauge-invariance $S_\kappa^g=S_\kappa$ is achieved provided the following constraints hold true 
\begin{eqnarray}
 J(\mathcal{E}^{d-2}\triangleright g)\star (\mathcal{E}\triangleright g^\dag)&=&\bbone\label{contraint1},\\
(\mathcal{E}^{-1}\triangleright g) \star J(\mathcal{E}\triangleright g^\dag)&=&\bbone\label{contraint2},
\end{eqnarray}
for any $g\in\mathcal{U}$. The last constraint \eqref{contraint2} is verified provided 
\begin{equation}
J=\mathcal{E}^{-2}\label{eqn1bis}
\end{equation}  
owing to \eqref{hopf1bis} and $g\star g^\dag=1$. Then, plugging \eqref{eqn1bis} into \eqref{contraint1} yields 
\begin{equation}
\mathcal{E}^{d-4}(g)\star\mathcal{E}(g^\dag)=\bbone,
\end{equation}
which fixes the unique allowed value for  the classical dimension of the $\kappa$-Minkowski space in which the gauge invariance can occur, namely 
\begin{equation}
d=5.
\end{equation}

Let us summarize the above analysis. Given the differential calculus related to $\mathfrak{D}_0$, a gauge-invariant action of the form \eqref{action1}, the natural noncommutative analog of a gauge theory action exists only in 5-dimensional $\kappa$-Minkowski space, i.e. one spatial dimension more than one could have expected. It takes the form
\begin{equation}
S^{\mathfrak{D}_0}_\kappa=\int d^5x \ F^\dag_{\mu\nu}\star (\mathcal{E}^{-2}\triangleright  F_{\mu\nu})\label{action1bis},
\end{equation}
invariant under the gauge transformation \eqref{fg1}.\\

At this point, some important comments are in order:\\

\begin{enumerate}
\item We point out that the derivations \eqref{modularDirac} are the only twisted derivations in $\mathcal{T}_\kappa$ leading to a (self-adjoint) Dirac operator $\mathcal{D}:=\Gamma^\mu X_\mu$ ($\Gamma^\mu$ being gamma matrices), having the correct commutative limit and from which one can construct a (necessarily twisted) spectral triple modeling a spectral geometry related to the $\kappa$-Minkowski space. This is the Theorem 20 of ref \cite{matas}. \\
Indeed, twisting the spectral triple must be achieved to ensure that one can actually define a commutator of the type $[\mathcal{D},\pi(a)]$ which is a bounded operator, $\pi(a)$ being some suitable representation of the algebra. This is an essential property for spectral triples. It appears that boundedness cannot be obtained with naive Dirac operator of the form $\Gamma^\mu P_\mu$ and usual (untwisted) commutator (say $[a,b]=ab-ba$). Twisting the commutator by the modular twist (say $[a,b]_\sigma=ab-\sigma(b)a$) and replacing the naive Dirac operator by $\mathcal{D}$ given above reaches the goal.{\footnote{One has $[\mathcal{D},\pi(a)]_\sigma\in\mathcal{B}(\mathcal{H})$ where the Hilbert space $\mathcal{H}$ and the representation $\pi$ of the (smooth) algebra for $\kappa$-Minkowski $\mathcal{M}_\kappa^0$ are obtained through a standard GNS construction.}}

\item This observation singles out the twisted derivations \eqref{modularDirac} together with the related differential calculus and twisted connection as a very natural framework able to capture important features of the noncommutative geometry of $\kappa$-Minkowski space. In particular, one salient physical prediction emerges within this framework: Assuming $\kappa$ of the order of the Planck mass, the coexistence of the $\kappa$-Poincar\'e invariance and the gauge invariance at the Planck scale predicts/favors the existence of one additional spatial dimension.\\
As far as the commutative limit of \eqref{action1bis} is concerned, note that a particular mechanism should be used to get rid of one extra spatial dimension, e.g. some compactification on $\mathbb{S}^1$, and combined to the commutative limit in order to obtain a 4-dimensional commutative (low energy effective) theory while taking the commutative limit alone in \eqref{action1bis} yields obviously a standard action of an Abelian gauge theory in five dimensions.\\
\end{enumerate}

The above analysis can be extended to general Abelian Lie algebra of bitwisted derivations \eqref{Leibniz-sigtau} as we now show.

\subsection{Extension to bitwisted connections.}\label{section44}

It is instructive to extend the notion of twisted connection elaborated in the subsection \ref{section32} to the case of a general  Abelian Lie algebra of bitwisted derivations with $\alpha\ne\bbone$, such as the one \eqref{sigtau-famil} generating $\mathfrak{D}_\gamma$, $\gamma\ne0$. Note by the way that the ensuing construction formally applies to any Lie algebra $\mathfrak{D}$ whose elements are not necessarily in $\mathcal{T}_\kappa$ (while one has $\mathfrak{D}_\gamma\subset\mathcal{T}_\kappa$).\\

Still assuming that $\mathbb{E}$ is a copy of $\mathcal{M}_\kappa^d$ and that we start from a differential calculus generated by an Abelian Lie algebra of bitwisted derivations with arbitrary twists $\alpha$ and $\beta$\footnote{Strictly speaking, $\alpha$ and $\beta$ should depend on the index $\mu$ of the derivation $X_{\mu}$ they are associated with, but we can show that they actually don't.}, the relevant bitwisted connection can be defined as a map $\nabla_{X_{\mu}}:\mathbb{E}\to\mathbb{E}$ with \eqref{sigtaucon3} changed into
\begin{equation}
\nabla_{X_{\mu}}(m\star a)=\nabla_{X_{\mu}}(m)\star\alpha(a)+\beta(m)\star X_{\mu}(a),\label{sigtauconbis}
\end{equation}
while \eqref{sigtaucon1} and \eqref{sigtaucon2} are unchanged and one has 
\begin{equation}
\nabla_{X_{\mu}}(a)=A_\mu\star\alpha(a)+X_{\mu}(a),\label{sigtauconter}
\end{equation}
for any $m\in\mathbb{E}$, $a\in\mathcal{M}_\kappa^d$, where we have set $A_\mu:=\nabla_{X_{\mu}}(\bbone)$\footnote{Here and from now on, we drop the factor $i$ that was introduced previously.}.\\

The related twisted gauge transformations are defined as
\begin{equation}
\nabla_{X_{\mu}}^g(a)=\beta\circ\rho(g^\dag)\star\nabla_{X_{\mu}}(\rho(g)\star a)\label{gaugebitwist},
\end{equation}
for any $g\in\mathcal{U}$, $a\in\mathcal{M}_\kappa^d$, where now the gauge group acts in a twisted way on the algebra, as $a^g=\rho(g)\star a$. Therefore, we allow that the ``matter'' gauge transformation \eqref{gaugematter} to be twisted by some automorphism $\rho$. The resulting gauge transformation on the ``gauge potential'' $A_\mu$ becomes
\begin{equation}
A_\mu^g=\beta\circ\rho(g^\dag)\star A_\mu\star\alpha\circ\rho(g)+\beta\circ\rho(g^\dag)\star X_\mu(\rho(g))\label{amugtwist}
\end{equation}
for any $g\in\mathcal{U}$, with $\nabla_{X_{\mu}}^g(a)=A_\mu^g\star\alpha(a)+X_{\mu}(a)$.\\

It is convenient to define
\begin{equation}
F_{\mu\nu}=X_\mu(\beta^{-1}(A_\nu))-X_\nu(\beta^{-1}(A_\mu))+A_\mu\star\alpha\circ\beta^{-1}(A_\nu)-A_\nu\star\alpha\circ\beta^{-1}(A_\mu)\label{bitwistcurv},
\end{equation}
stemming from\footnote{Note that, as a consequence, $F$ is not a morphism of module like in the commutative case. It is a morphism of twisted module.}
\begin{equation}
\nabla_{X_{\mu}}(\beta^{-1}(\nabla_{X_{\nu}}(\bbone \star a)))-\nabla_{X_{\nu}}(\beta^{-1}(\nabla_{X_{\mu}}(\bbone \star a)))=F_{\mu\nu}\star\alpha^2\circ\beta^{-1}(a) \label{curvature-real}.
\end{equation}
From the gauge transformation \eqref{amugtwist}, one obtains after some algebra
\begin {equation}
F_{\mu\nu}^g=\beta\circ\rho(g^\dag)\star F_{\mu\nu}\star\alpha^2\circ\beta^{-1}\circ\rho(g),\ \ \forall g\in\mathcal{U}.\label{decadix2}
\end{equation}
Then, the requirement of the gauge invariance of the action
\begin{equation}
S_\kappa=\int d^dx \ F^\dag_{\mu\nu}\star J(F_{\mu\nu})\label{action22},
\end{equation}
where again $J(a\star b)=J(a)\star J(b)$ as in the subsection \ref{section43}, leads to
\begin{eqnarray}
\mathcal{E}^{d-1}\circ J\circ\alpha^2\circ\beta^{-1}\circ\rho(g)\star\alpha^{-2}\circ\beta\circ\rho^{-1}(g^\dag)=1\label{decadix31}\\
\beta^{-1}\circ\rho^{-1}(g)\star J\circ\beta\circ\rho(g^\dag)=1\label{decadix32}.
\end{eqnarray}
Eqn. \eqref{decadix32} implies
\begin{equation}
\beta\circ\rho=J^{-\frac{1}{2}}\label{betarho},
\end{equation}
which combined with \eqref{decadix31} yields
\begin{equation}
\beta\circ\alpha^{-1}=\mathcal{E}^{\frac{d-1}{4}}.\label{relat-critik}
\end{equation}
Let us discuss these results:\\

\begin{enumerate}
\item Assume that the ``matter'' gauge transformations are untwisted, i.e. $\rho=\bbone$. Then, choosing $\mathfrak{D}_\gamma$ for which $\alpha=\mathcal{E}^\gamma$ and $\beta=\mathcal{E}^{\gamma+1}$, one observes that \eqref{relat-critik} selects the unique value $d=5$, valid for any $\gamma\in\mathbb{R}$, for which the gauge invariance holds while \eqref{betarho} gives $J=\mathcal{E}^{-2(\gamma+1)}$. The resulting gauge invariant action is 
\begin{equation}
S_\kappa=\int d^5x \ F^\dag_{\mu\nu}\star (\mathcal{E}^{-2(\gamma+1)}\triangleright F_{\mu\nu}),
\end{equation}
while the corresponding gauge transformations can be readily obtained from \eqref{bitwistcurv}. Note that twisting the ``matter'' gauge transformations eqn. \eqref{gaugematter} by using $\rho\ne\bbone$ does not change the value for the classical dimension d but only affects $J$.
\item Eqn. \eqref{relat-critik} exhibits a rigid relation between gauge-invariance of the action \eqref{action22}, the classical dimension of the $\kappa$-Minkowski space and the twist properties characterizing the twisted differential calculus.  In the general case of bitwisted differential calculi presented in the subsection \ref{section31}, one observes from \eqref{relat-critik} that the choice of a differential calculus (i.e. choosing $\alpha$ and $\beta$) actually fixes to a unique value the dimension of the $\kappa$-Minkowski space for which the gauge invariance of the actions of the form \eqref{action22} can exist.
\end{enumerate}

\section{Conclusion.}

\noindent In this paper, we have discussed this problem within various large classes of untwisted and (bi)twisted differential calculi and finally we have provided an explicit solution of physical interest, starting from a natural class of noncommutative differential calculi based on (bi)twisted derivations of $\mathcal{T}_\kappa$ combined with a twisted extension of the notion of connection. \\
Namely, looking for a reasonable $\kappa$-Poincar\'e invariant analog of gauge invariant actions in terms of polynomials in the curvature, we have established an algebraic relation between the various twists and the classical dimension of the $\kappa$-Minkowski space which ensures the gauge invariance of the candidate actions. Fixing the twists fixes the unique value of the dimension at which the gauge invariance can be achieved while fixing the dimension severely restricts the allowed twists.\\
Besides, we have shown that within standard (untwisted) differential calculi, such as those usually considered in the physics literature, there is no (non trivial) polynomial actions in the curvature having the full gauge invariance; however such actions still remain invariant under a group of transformations constrained by the Tomita operator stemming from the $\kappa$-Poincar\'e invariance. \\
Among the above class of (bi)twisted derivations, there is a distinguished unique set of derivations leading to a Dirac operator with required properties to be used in a (twisted) spectral triple modeling $\kappa$-Minkowski space. Using this unique set in $\mathcal{T}_\kappa$ singles out d=5 as the unique dimension for which the above gauge actions can support both the gauge invariance and the $\kappa$-Poincar\'e invariance.\\

It could be interesting to study the perturbative behavior of the 5-dimensional gauge invariant action \eqref{action1bis} as well as to use some mechanism (e.g. compactification) to get rid of one extra spatial dimension and examine the resulting model. The extension of the present work to the case of (linear) connections on a bimodule may of course exhibit additional interesting features leaving some room for the appearance of curvature and possibly torsion in the corresponding gauge invariant actions. Such an analysis has been undertaken \cite{mw-1}. According to \eqref{relat-critik}, reconciling both the $\kappa$-Poincar\'e invariance and the gauge invariance in a 4-dimensional $\kappa$-Minkowski space would be achieved starting from bitwisted derivations satisfying $\beta\alpha^{-1}=\mathcal{E}^{\frac{3}{4}}$ which therefore cannot belong to $\mathcal{T}_\kappa$ as shown in subsection \ref{section31}. Hence, within the framework developed in this paper, the $\kappa$-Poincar\'e invariance and the gauge invariance can be reconciled only in a five dimensional $\kappa$-Minkowski space-time.

\vskip 2 true cm
{\bf{Acknowledgments:}} One of us (J.-C. W.) thanks P. Martinetti for numerous exchanges on the use of twists in almost noncommutative geometry and F. Besnard and M. Dubois-Violette for discussions at various stages of this work.

\appendix
\section{Twisted convolutions and Weyl-Wigner quantization.}\label{apendmoyal}

It has been known for a long time that the Weyl-Wigner quantization scheme gives rise to the Moyal product. This latter is nothing but the inverse Fourier transform of a twisted convolution product \cite{jvn} stemming from the convolution product equipping $L^1(\mathbb{H})$, the convolution algebra of the unimodular Heisenberg group $\mathbb{H}$. \\

Sketching the construction of the Moyal product in the 2-dimensional case (the extension to 4 dimensions is straightforward), one first observes that the functions $f\in L^1(\mathbb{H})$ can be viewed as functions $f(z;p,q)$, hence as functions in $L^1(\mathbb{R}^3)$. Here, $p,q,z$ generate the 3-d Heisenberg Lie algebra $[p,q]=iz$, and $z$ is a central element. \\
Next, one realizes that (nondegenerate) representations of $L^1(\mathbb{H})$, $\pi:L^1(\mathbb{H})\to\mathcal{B}(L^2(\mathbb{R}))${\footnote{$\mathcal{B}(L^2(\mathbb{R}))$ denotes as usual the algebra of bounded operator in $L^2(\mathbb{R})$ .}} can be expressed as 
\begin{equation}
(\pi(f)\psi)(x)=\int_{\mathbb{R}^2}dpdq\ f^\#(p,q)e^{i(qx+h\frac{pq}{2}})\psi(x+hp),\ h\ne0, 
\end{equation}
owing to the Stone-von Neumann theorem, where 
\begin{equation}
f^\#(p,q)=\int dz e^{ihz}f(p,q,z)
\end{equation}
defines a map $\#:L^1(\mathbb{R}^3)\to  L^1(\mathbb{R}^2)$. \\
By a standard computation, one further observes that 
\begin{equation}
(f\odot g)^\#:=f^\#\hat{\odot}g^\#, 
\end{equation}
in which $\odot$ denotes the usual convolution product equipping $L^1(\mathbb{H})$ while ${\hat{\odot}}$ is the so-called twisted convolution product.\\
 Finally, by identifying the functions $f^\#(p,q)$ as Fourier transforms, namely $f^\#(p,q)=(\mathcal{F}f)(p,q)$ and using the Weyl quantization map defined by
\begin{equation}
W(f):=\pi(\mathcal{F}f),
\end{equation}
with 
\begin{equation}
W(f\star_M g)=W(f)W(g), 
\end{equation}
one obtains the usual expression for the Moyal product $\star_M$ given by
\begin{equation}
f\star_M g=\mathcal{F}^{-1}(\mathcal{F}f\hat{\odot}\mathcal{F}g).
\end{equation}

\section{4-d $\kappa$-Poincar\'e Hopf algebra.}\label{apendixA}

We denote $\mathcal{P}_\kappa$ the $\kappa$-Poincar\'e algebra and $\Delta:\mathcal{P}_\kappa\to\mathcal{P}_\kappa\otimes\mathcal{P}_\kappa$, $\epsilon:\mathcal{P}_\kappa\to\mathbb{C}$ and $S:\mathcal{P}_\kappa\to\mathcal{P}_\kappa$ respectively the coproduct, counit and antipode. Recall that $(\mathcal{P}_\kappa,\ \Delta,\ \epsilon,\ S)$ define a Hopf algebra. We consider the case of the classical dimension equal to 4. The extension to the d-dimensional case is obvious.\\
A presentation of $\mathcal{P}_\kappa$ can be obtained from the 11 elements $(P_i, N_i,M_i, \mathcal{E},\mathcal{E}^{-1})$, $i=1,2,3$, respectively the momenta, boosts, rotations and $\mathcal{E}:=e^{-P_0/\kappa}$ satisfying the Lie algebra relations{\footnote{Greek (resp. Latin) indices label space-time (resp. purely spatial) coordinates.}}
\begin{equation}
[M_i,M_j]= i\epsilon_{ij}^{\hspace{5pt}k}M_k,\ [M_i,N_j]=i\epsilon_{ij}^{\hspace{5pt}k}N_k,\ [N_i,N_j]=-i\epsilon_{ij}^{\hspace{5pt}k}M_k\label{poinc1}, 
\end{equation}
\begin{equation}
[M_i,P_j]= i\epsilon_{ij}^{\hspace{5pt}k}P_k,\ [P_i,\mathcal{E}]=[M_i,\mathcal{E}]=0,\ [N_i,\mathcal{E}]=-\frac{i}{\kappa}P_i\mathcal{E}\label{poinc2},
\end{equation}
\begin{equation}
[N_i,P_j]=-\frac{i}{2}\delta_{ij}\left(\kappa(1-\mathcal{E}^{2})+\frac{1}{\kappa}\vec{P}^2\right)\label{poinc3}.
\end{equation}
The relations defining the Hopf algebra structure are
\begin{align}
\Delta P_0&=P_0\otimes\bbone+\bbone\otimes P_0,\ \Delta P_i=P_i\otimes\bbone+\mathcal{E}\otimes P_i,\label{hopf1}\\
\Delta \mathcal{E}&=\mathcal{E}\otimes\mathcal{E},\ \Delta M_i=M_i\otimes\bbone+\bbone\otimes M_i,\label{hopf1bis}\\
\Delta N_i&=N_i\otimes \bbone+\mathcal{E}\otimes N_i-\frac{1}{\kappa}\epsilon_{i}^{\hspace{2pt}jk}P_j\otimes M_k,\label{hopf2}\\
\epsilon(P_0)&=\epsilon(P_i)=\epsilon(M_i)=\epsilon(N_i)=0,\  \epsilon(\mathcal{E})=1\label{hopf3},\\
S(P_0)&=-P_0,\ S(\mathcal{E})=\mathcal{E}^{-1},\  S(P_i)=-\mathcal{E}^{-1}P_i,\  S(M_i)=-M_i,\label{hopf4}\\
S(M_i)&=-M_i,\ S(N_i)=-\mathcal{E}^{-1}(N_i-\frac{1}{\kappa}\epsilon_{i}^{\hspace{2pt}jk}P_jM_k)\label{hopf4bis}.
\end{align}
The $\kappa$-Minkowski space can be described as the dual of the Hopf subalgebra generated by the $P_\mu$'s and $\mathcal{E}$. Let $\mathcal{T}_\kappa$ denote this algebra, called the ``deformed translation algebra''. This latter can be equipped with an involution, hence becoming a $^*$-Hopf algebra, through $P_\mu^\dag=P_\mu$, $\mathcal{E}^\dag=\mathcal{E}$. The extension of the above duality to a duality between involutive algebras is achieved through
\begin{equation}
(t\triangleright f)^\dag=S(t)^\dag\triangleright f,\label{pairing-involution}
\end{equation}
which holds true for any $t\in\mathcal{T}_\kappa$ and any $f\in\mathcal{M}_\kappa^4$. From \eqref{pairing-involution} and \eqref{hopf4}, one easily obtains
\begin{equation}
(P_0\triangleright f)^\dag=-P_0\triangleright(f^\dag),\ (P_i\triangleright f)^\dag=-\mathcal{E}^{-1}P_i\triangleright(f^\dag),\ (\mathcal{E}\triangleright f)^\dag=\mathcal{E}^{-1}\triangleright(f^\dag)\label{dag-hopfoperat}.
\end{equation}
As far as the action of $\mathcal{T}_\kappa$ on $\mathcal{M}_\kappa^4$ is concerned, recall that the $P_i$'s act as twisted derivations while $P_0$ is a standard derivation. Indeed from \eqref{hopf1}, one can write for any $f,g\in\mathcal{M}_\kappa^4$
\begin{align}
P_i\triangleright(f\star g)&=(P_i\triangleright f)\star g+(\mathcal{E}\triangleright f)\star (P_i\triangleright g)\label{deriv-twist1},\\
P_0\triangleright(f\star g)&=(P_0\triangleright f)\star g+f\star(P_0\triangleright  g )\label{deriv-twist2}.
\end{align}
$\mathcal{E}$ is simply an automorphism of $\mathcal{M}_\kappa^4$ (not a derivation) and one has
\begin{equation}
\mathcal{E}\triangleright(f\star g)=(\mathcal{E}\triangleright f)\star(\mathcal{E}\triangleright g).\label{relation-calE}
\end{equation}

Recall that $\mathcal{M}_\kappa^4$ is a left-module over the Hopf algebra $\mathcal{P}_\kappa$ and one has, for any $f\in\mathcal{M}_\kappa^4$, in the so-called bicrossproduct basis $(M_i,N_i,P_\mu)$ 
\begin{align}
(\mathcal{E}\triangleright f)(x)&=f(x_0+\frac{i}{\kappa},\vec{x})\label{left-module0},\\
(P_\mu\triangleright f)(x)&=-i(\partial_\mu f)(x),\\ 
(M_i\triangleright f)(x)&=\left(\epsilon_{ij}^{\hspace{5pt}k}x^jP_kf\right)(x),\label{left-modules1bis}\\
(N_i\triangleright f)(x)&=\left(\left[\frac{\kappa}{2}L_{x_i}(\mathcal{E}-\mathcal{E}^{-1})+L_{x_0}P_i\mathcal{E}+L_{x_i}\vec{P}^2\mathcal{E}\right]f\right)(x),\label{left-module2}
\end{align}
in which $L_a$ denotes the left (standard) multiplication operator, i.e. $L_af:=af$. \\

\section{Hermitian structure and gauge transformations.}\label{apend-gaugegroup}
Let $\mathbb{E}$ be a right-module over $\mathcal{M}_\kappa^d$ and $h$ a Hermitian structure, that is a sesquilinear form $h:\mathbb{E}\otimes\mathbb{E}\to\mathcal{M}_\kappa^d$ satisfying:
\begin{equation}
h(m_1,m_2)^\dag=h(m_2,m_1), \ \ \ h(m_1\bullet a_1,m_2\bullet a_2)=a_1^\dag\star h(m_1,m_2)\star a_2\label{hermit-struc-cond}, 
\end{equation}
for any $m_1,m_2\in\mathbb{E}$ and any $a_1,a_2\in\mathcal{M}_\kappa^d$ where $m\bullet a$ denotes the action of the algebra on the module. \\
We assume that $\mathbb{E}$ is a copy of the algebra $\mathcal{M}_\kappa^d$ and the Hermitian structure is
\begin{equation}
h_0(m_1,m_2)=m_1^\dag \star m_2, \label{hermit-structure}
\end{equation}
for any $m_1,m_2\in\mathbb{E}$. The action of the algebra on $\mathbb{E}$ is simply given by 
\begin{equation}
m\bullet a=m\star a, \label{module-action}
\end{equation}
which obviously fulfills \eqref{hermit-struc-cond}. \\
Untwisted gauge transformations are defined as the set of automorphisms of $\mathbb{E}$ preserving its right-module structure on $\mathcal{M}_\kappa^d$ and compatible with the Hermitian structure, given here by $h_0$. Let us denote this set $\text{Aut}_{h_0}(\mathbb{E})$. For any $\varphi\in\text{Aut}_{h_0}(\mathbb{E})$, one therefore must have
\begin{equation}
h_0(\varphi(m_1),\varphi(m_2))=h_0(m_1,m_2)\label{compatibility-h}
\end{equation}
which holds true for any $m_1,m_2\in\mathbb{E}$. Since $\varphi(m\star a)=\varphi(m)\star a$ for any $m\in\mathbb{E}$, $a\in\mathcal{M}_\kappa^d$ as a morphism of module, one can write $\varphi(\bbone\star a)=\varphi(\bbone)\star a$ (keeping in mind that here $\bbone\in\mathbb{E}$) so that the action of any gauge transformation $\varphi$ on the algebra $\mathcal{M}_\kappa^d$ is entirely determined by its action on the unit. Accordingly, we define
\begin{equation}
a^g:=\varphi(a)=g\star a, \ \ \varphi(\bbone):=g.\label{gaugematter}
\end{equation}
Then, by simply writing $h_0(\varphi(m_1\star a_1),\varphi(m_2\star a_2))$ for $m_1=m_2=\bbone$, the requirement \eqref{compatibility-h} of compatibility of the Hermitian structure with gauge transformations yields
\begin{equation}
g^\dag\star g=g\star g^\dag=\bbone, \label{unitar-gauge}
\end{equation}
which thus defines the noncommutative analog of unitary gauge transformations. Accordingly, we use the following convenient definition of the gauge group:
\begin{equation}
\mathcal{U}:=\{g\in\mathbb{E},\ \ g^\dag\star g=g\star g^\dag=\bbone \}\label{unitar-group},
\end{equation}
which completely characterizes $\text{Aut}_{h_0}(\mathbb{E})$.

%-- BIBLIOGRAPHY ------------------------------------------------------------%
{\small%

}%
\end{document}